\documentclass[aps,prl,preprint]{revtex4}
\usepackage{graphicx}
\begin{document}
\title{Observation of quantum capacitance in the Cooper-pair transistor}
\author{T.~Duty}
\email[]{tim@mc2.chalmers.se}
\author{G.~Johansson}
\author{K.~Bladh}
\author{D.~Gunnarsson}
\author{C.~Wilson}
\author{P.~Delsing}
\affiliation{Microtechnology and Nanoscience, MC2, Chalmers
University of Technology, S-412 96 G{\"o}teborg, Sweden}
\date{\today}
\begin{abstract}

We have fabricated a Cooper-pair transistor (CPT) with parameters
such that for appropriate voltage biases, the sub-gap charge
transport takes place via slow tunneling of quasiparticles that
link two Josephson-coupled charge manifolds.  In between the
quasiparticle tunneling events, the CPT behaves essentially like a
single Cooper-pair box (SCB). The effective capacitance of a SCB
can be defined as the derivative of the induced charge with
respect to gate voltage. This capacitance has two parts, the
geometric capacitance, $C_\mathrm{geom}$, and the quantum
capacitance $C_Q$. The latter is due to the level anti-crossing
caused by the Josephson coupling. It depends parametrically on the
gate voltage and is dual to the Josephson inductance. Furthermore,
it's magnitude may be substantially larger than $C_\mathrm{geom}$.
We have been able to detect $C_Q$ in our CPT, by measuring the
in-phase and quadrature rf-signal reflected from a resonant
circuit in which the CPT is embedded. $C_Q$ can be used as the
basis of a charge qubit readout by placing a Cooper-pair box in
such a resonant circuit.
\end{abstract}
%\pacs{74.50.+r,85.25.Cp,03.67.Lx,03.65.-w}
\maketitle
\graphicspath{{..}}

The quantum-mechanical properties of the single Cooper-pair box
(SCB)\cite{buettiker87,bouchiat98}---an artificial two-level
system--have been investigated thoroughly during the last few
years due to the potential for SCB's to serve as quantum bits
(qubits)\cite{nakamura99,makhlin01,vion02,lehnert03,duty_prb04}.
An important property of a so-called charge qubit is the existence
of an optimal point, where the first derivative of the energy
bands with respect to gate voltage vanishes, and the system is
insensitive to low-frequency charge fluctuations\cite{vion02,
duty_prb04}. Dephasing times are maximum at this point, making it
the natural operation point for single-qubit quantum rotations.
Since the eigenstates at the optimal point are orthogonal to
charge eigenstates, however, one must move away from the optimal
point for read-out schemes based upon charge measurement.

A recent experiment used the polarizability of an SCB coupled to a
microwave resonator to perform cavity-QED
measurements\cite{wallraff04}. Such a circuit can also perform a
quantum non-demolition measurement of the qubit state at the
optimal point. In this paper, we study a type of polarizability
that can be described as an effective capacitance and is related
to the second derivative, or curvature, of the energy bands with
respect to gate voltage. This quantum capacitance was first
discussed in the context of small Josephson
junctions\cite{widom84,averin85,likharev85} and is dual to the
Josephson inductance. Recently, a controllable coupling scheme
based upon this parametric capacitance\cite{avebru03} has been
proposed, as well as a superconducting phase
detector\cite{roschier04}.

Both the single-electron transistor
(SET)\cite{fulton87,likharev87} and its superconducting version,
also known as the Cooper-pair transistor (CPT)\cite{zorin96}, are
closely related to the SCB. These devices are the basis of very
sensitive electrometers that are used to read-out charge
qubits\cite{aassime01b,lehnert03,duty_prb04}. Previous work
concentrated on the dissipative response and back-action of SET's
and CPT's when used as
electrometers\cite{aassime01,johansson02,choi01,clerk02}. Here, we
show that an appropriately designed CPT can also exhibit a
reactive response due to quantum capacitance. This capacitance can
be measured using a radio-frequency resonant circuit.

\begin{figure}[tb]
    \centering
    \includegraphics[width =0.7\columnwidth]{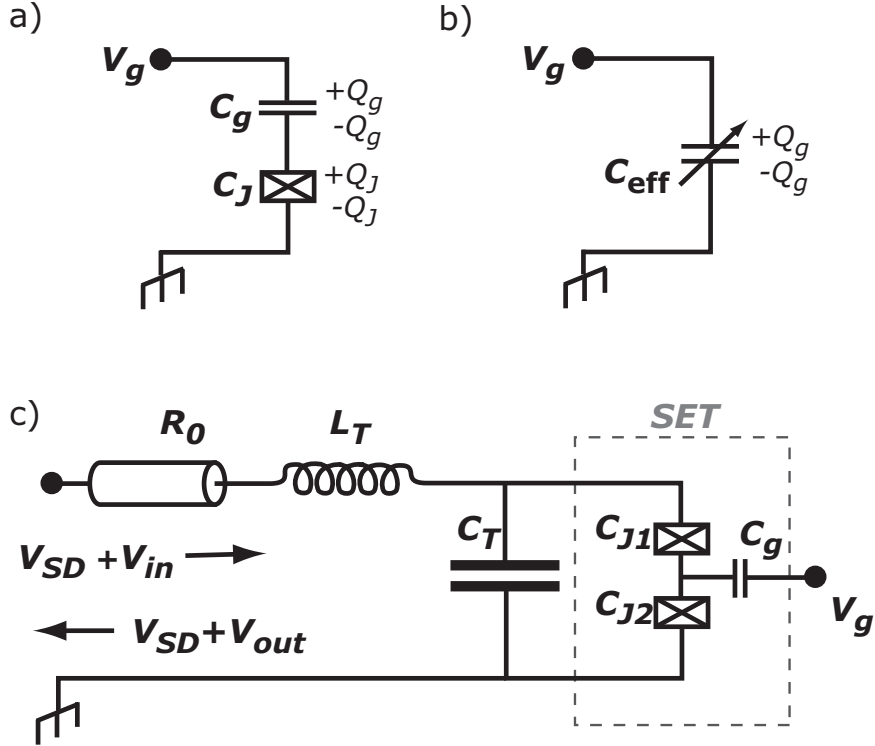}
    \caption{a) The Cooper-pair box and b) it's equivalent as a parametric capacitor.
             c) schematic of the RF-SET. A monochromatic radio-frequency signal
             $V_\mathrm{in}$, added to a DC voltage bias $V_\mathrm{SD}$,
             is reflected from a tank circuit containing the SET.}
    \label{schematics}
\end{figure}
To see how quantum capacitance arises, we first consider the
single Cooper-pair box as depicted in Fig. \ref{schematics}a. The
box has a Josephson energy $E_J$, charging energy
$E_C=e^2/2C_{\Sigma}$ and total capacitance $C_{\Sigma}=C_J+C_g$.
The effective capacitance can be defined as the first derivative
of the injected charge with respect to voltage,
$C_\mathrm{eff}=\partial \left<Q_g\right>/\partial V_g$,
%\begin{equation}\label{ceff_cpb}
%    C_\mathrm{eff}=\frac{\partial \left<Q_g\right>}{\partial V},
%\end{equation}
where the brackets denote a quantum expectation value. From
electrostatics, one has $ Q_g=C_g(V_g-V_\mathrm{island})$, and
$V_\mathrm{island}=(C_gV_g-2en)/C_\Sigma$, so that
\begin{equation}\label{qg_cpb}
    \left<Q_g\right>=\frac{C_gC_J}{C_\Sigma}V_g+2e\left<n\right>\frac{C_g}{C_\Sigma},
\end{equation}
where $n$ is the number of Cooper pairs that have tunneled onto
the island. For each energy band $k$ of the SCB, $\left<n\right>$
depends on the normalized gate charge $n_g=C_gV_g/e$. One finds
that
\begin{eqnarray}
  C^k_{\mathrm{eff}} = \frac{C_gC_J}{C_\Sigma}- \frac{C_g^2}{e^2}
 \frac{\partial^2 E_k}{ \partial
    n_g^2}
   = C_\mathrm{geom}-C^k_Q
\end{eqnarray}
where $C_\mathrm{geom}$ is the \emph{geometric} capacitance of two
capacitors in series, and $C^k_Q\equiv (C_g^2/e^2)(\partial^2 E_k/
\partial n_g^2)$ is the \emph{quantum} capacitance. In the two-level
approximation, which is valid for $\epsilon \equiv E_J/4E_C\ll 1$,
the ground and first excited state energies are given by
$E_\pm=\pm2E_C\sqrt{(1-n_g)^2+\epsilon^2}$. These produce the
quantum capacitances
%\begin{equation}\label{cq_two}
%    C_Q^\pm=\pm \frac{C_g^2}{C_{\Sigma,b}}
%    \frac{\epsilon^2}{\left( \left(1-n_g\right)^2+ \epsilon^2 \right)^{3/2}}.
%\end{equation}
\begin{equation}\label{cq_two}
    C_Q^\pm=\pm \frac{C_g^2}{C_\Sigma}
    \epsilon^2\left( \left(1-n_g\right)^2+ \epsilon^2 \right)^{-3/2}.
\end{equation}
The magnitude of $C_Q$ is maximum at the charge degeneracy ,
\begin{equation}\label{cd_qcap}
    C_Q^\pm(n_g=1)=\pm\frac{2e^2}{E_J}\frac{C_g^2}{C_\Sigma^2}.
\end{equation}
and $C_Q$ is negative in the ground state. We note that although
$C_Q$ grows with decreasing $E_J$, the region of $n_g$ where it is
observable becomes vanishingly small. The value of $e^2/h$ is
approximately $40$ fF$\cdot$GHz, so a charge qubit (SCB) with
$E_J/h=10$ GHz and $C_g=C_J$ would have a quantum capacitance at
the charge degeneracy of 2 fF, which is higher than the typical
junction capacitance ($C_J\sim$ 1fF) of a charge qubit.

For finite temperatures, a thermal expectation of the injected
charge must be considered, giving one the effective capacitance
\begin{eqnarray} \label{thermal_cq}
  C_\mathrm{eff} = \frac{C_gC_J}{C_\Sigma}- \frac{C_g^2}{e^2}\frac{\partial}{ \partial
    n_g}\left< \frac{\partial E_k}{ \partial
    n_g}\right>_\mathrm{T},
\end{eqnarray}
where $\left<...\right>_\mathrm{T}$ denotes a Boltzmann-weighted
average over the energy bands.

\begin{figure}[tb]
    \centering
    \includegraphics[width = 0.9\columnwidth]{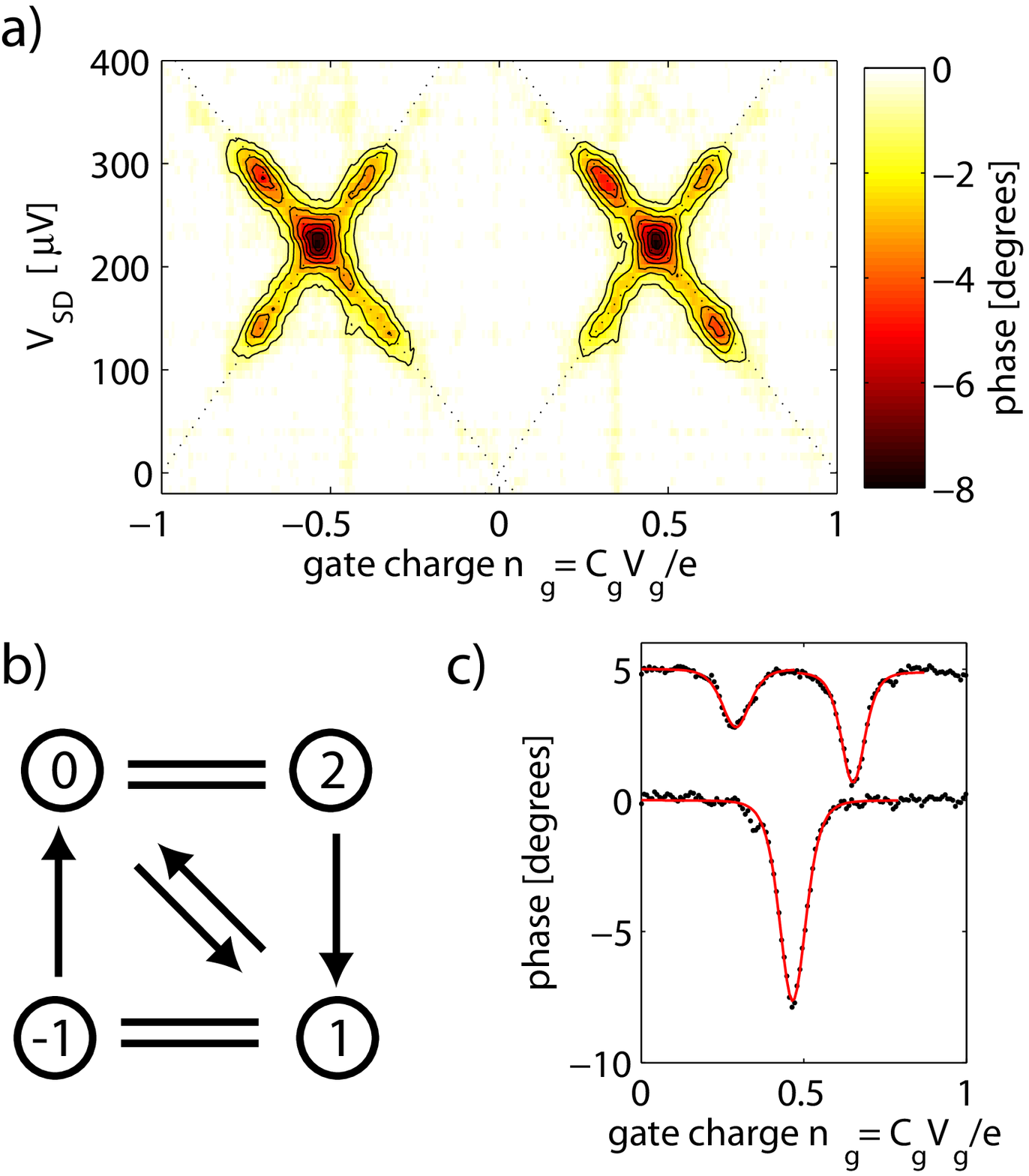}
    \caption{a) Phase shifts around $eV_\mathrm{SD}=2E_C$. Cooper-pair tunneling
    across junction 1 (2) is degenerate along the the dotted lines with
    positive (negative) slope.  b) diagram of the
    charge states involved in the JQP processes for gate charge
    $0<n_g<1$. The solid lines with arrows represent tunneling of
    quasiparticles and the double lines Cooper pairs. The upper right
    triangle represents transitions
    occurring along the negative-slope dotted line in (a), and
    the bottom left triangle those along the positive-slope line.
    c). Phase shifts vs. gate charge for $V_\mathrm{SD}=222\mu V$ (lower points) and
    $V_\mathrm{SD}=145\mu V$ (upper points). The upper data are offset 5 degrees
    for clarity. The solid lines are the theoretical expectations using
    Eq. \ref{thermal_cq} and the two-level approximation, the values
    of $E_J$ found from spectroscopy, and assuming a temperature of 130 mK.}
  \label{phase}
\end{figure}

We now turn to the CPT, which consists of a metallic island
connected to two leads by small capacitance tunnel junctions (see
Fig. \ref{schematics}c). An external gate controls the potential
of the island through the gate-induced charge on the island,
$n_g=C_gV_g/e$. If $E_J/4E_C\ll 1$, an appreciable direct
quasiparticle (QP) current occurs only when
$eV_\mathrm{SD}>4\Delta$. For smaller bias voltages, a
gate-dependent sub-gap current is possible due to sequences of
Cooper-pair tunneling combined with QP tunneling. These processes
are known as Josephson-quasiparticle (JQP)
cycles\cite{fulton89,averin89,vandenbrink91a,vandenbrink91b,nakamura96}.
Recent research has focused upon describing the noise and
back-action effects of such JQP processes---when they carry a
substantial current\cite{choi01,clerk02}.
%\begin{figure}[tb]
%    \centering
%    \includegraphics[width =1.0\columnwidth]{schematic.eps}
%    \caption{Equivalent }
%    \label{schematic}
%\end{figure}
For a JQP process to carry a substantial current, however, the QP
tunneling rates must be relatively fast. We have constructed a CPT
where these rates are very slow in a certain region of voltage
bias $V_\mathrm{SD}$. This region of $V_\mathrm{SD}$ is centered
at $eV_\mathrm{SD}=2E_C$, at the intersections of lines in the
$V_\mathrm{SD}-n_g$ plane where Cooper-pair tunneling across one
junction is resonant (see Fig. \ref{phase}). This intersection is
known as the double JQP (DJQP) point.

Charge transport in this region of $V_\mathrm{SD}$ consists of QP
tunneling events that move the system between two
Josephson-coupled charge manifolds\cite{clerk02}. \textit{E.g.},
for $n_g \in [0,1]$, the QP transitions link the $0
\leftrightarrow 2$ and $-1 \leftrightarrow 1$ manifolds (see
Fig.\ref{phase}b). The QP tunneling rate $\Gamma_{qp}$ is related
to the QP current $I_{qp}$ of a single junction by
\begin{equation}\label{qprate}
\Gamma_{qp}(\Delta E)=\frac{I_{qp}(\Delta E)}{e} \frac{1}{exp
\left( -\Delta E/k_BT \right)-1},
\end{equation}
and depends on $\Delta E$, the gain in charging energy of the
tunnel event. Due to the superconducting density of states,
$I_{qp}(\Delta E)$ and hence $\Gamma_{qp}(\Delta E)$ is large only
when the energy gain exceeds $2\Delta$.
%For QP tunneling involving
%$\Delta E<0$, the small sub-gap $\Gamma_{qp}$ is further
%suppressed by the thermal factor in Eq. \ref{qprate}.
At the DJQP point,  $\Delta E=3E_C$, and hence QP tunneling is
suppressed if $E_C<2\Delta/3$ \cite{fulton89,nakamura96}, which is
the case for the CPT considered here, $E_C\simeq \Delta/2$. The
measured DC current at the DJQP point is less than $\sim 1$ pA,
which indicates a QP tunneling rate $\sim 4$ MHz. Because the QP
rates are very slow compared to the Cooper-pair tunneling rate
($\sim3$ GHz), and the intra-manifold relaxation rate ($\sim 1$
GHz, inferred from the spectroscopic measurements discussed
below), this CPT behaves essentially like a SCB, where Cooper
pairs tunnel coherently across one junction while the other
junction acts as a gate capacitance. This picture is interrupted
at long timescales by incoherent QP tunnel events.

The quantum states involved at these voltage biases are states of
definite number, both for the island charge, and the number of
charges having passed through the CPT. This description breaks
down in a small region around zero bias, $eV_\mathrm{SD}\ll E_C$,
due to the supercurrent, which can be thought of as arising from
the near degeneracy of states differing in the number of Cooper
pairs having passed through the CPT. The eigenstates at zero bias
are states of definite phase across the CPT. Since the effective
Josephson coupling depends on this phase, $C_Q$ at small but
finite $eV_\mathrm{SD}$ should be substantially reduced. At zero
bias, $C_Q$ is also sensitive to low-frequency fluctuations of the
phase. As a consequence, we do not observe $C_Q$ around zero bias,
which could also be due to poisoning by non-equilibrium QP's.

We fabricated a CPT using electron-beam lithography and standard
double-angle shadow evaporation of aluminum films onto an oxidized
silicon substrate. The sample was placed at the mixing chamber of
a dilution refrigerator with a base temperature of $\approx$ 20
mK. All DC control lines were filtered by a combination of
low-pass and stainless steel powder filters. The measured normal
resistance of the SET was $R_n=120k\Omega$ which implies a
Josephson energy $E_J=12\mu eV$ ($E_J/h$=2.9 GHz ) per junction
using the Ambegaokar-Baratoff relation\cite{ab63}. The charging
energy and superconducting gap were determined from the DC-IV
curves, $E_C=111$ $\mu$eV, and $\Delta=215$ $\mu$eV.

Our CPT was configured as a radio-frequency SET
(RF-SET)\cite{schoelkopf98}, which is based upon reflection of a
monocromatic radio-frequency signal from a tank (LC) circuit
containing the SET (see Fig. \ref{schematics}c). The tank circuit
described here had a resonant frequency of 342 MHz using a
inductance $L_T=490$ nH, which implies a tank circuit capacitance
$C_T$=440 fF coming from the stray capacitance of the bonding pad
connecting the inductor to the chip. The CPT has a reactive
response of the CPT due to it's effective capacitance. Like that
of the SCB, it is related to second derivatives of the energy
bands. In this case, the derivatives are with respect to the
source-drain voltage $V_\mathrm{SD}$, and one must tune both
$V_\mathrm{SD}$ and $n_g$ to sit at a Cooper-pair charge
degeneracy. One finds a form similar to Eq. \ref{cq_two} but with
$C_\Sigma=C_{J1}+C_{J2}+C_{g}$. For a symmetric transistor
($C_{J1}=C_{J2}$) and $C_g \ll C_{J1}$ the expression is identical
to Eq. \ref{cq_two}, but with $1-n_g$ replaced by $1-n_g-v/4$,
where $v=eV_\mathrm{SD}/E_C$.

The reflection coefficient for the circuit of Fig.
\ref{schematics}c is given by $\alpha
=V_\mathrm{out}/V_\mathrm{in}=(Z-Z_0)/(Z+Z_0)$,
%\begin{equation}\label{alpha}
%    \alpha = \frac{Z-Z_0}{Z+Z_0},
%\end{equation}
with
%\begin{equation}\label{tankimp}
%\quad Z= i \omega L_T + \frac{1}{i \omega
%    C+R_{SET}^{-1}},
%\end{equation}
\begin{equation}\label{tankimp}
\quad Z= i \omega L_T + \left(i \omega
    C+R_{SET}^{-1}\right)^{-1},
\end{equation}
where $C=C_T+C_\mathrm{eff}$ is the total capacitance, and
$Z_0\simeq50\Omega$. $C_\mathrm{eff}$ is the effective capacitance
of the CPT, which depends on both $V_\mathrm{SD}$ and $n_g$. If
$R_{SET} \gg L/Z_0C \simeq 23$ k$\Omega$ for our circuit, the
phase of $\alpha$ near resonance is only affected by changes in
$C$. It will be convenient to fix the total capacitance (and hence
phase) relative to some particular value of $V_\mathrm{SD}$ and
$n_g$, where $C_Q=0$. We write $C=C_0-C_Q$. We define the detuning
parameter $\delta=1-\omega/\omega_0$, with
$\omega_0=1/\sqrt{L_TC_0}$ and $\omega=1/\sqrt{L_TC}$. For
$2Q\delta \ll 1$, one finds $\alpha=-1+i4Q\delta$,
%\begin{equation}
%\label{a}
%  \alpha=-1+i4Q\delta,
%\end{equation}
where the quality factor $Q=\sqrt{L_T/C}/Z_0=$21 for our circuit,
and $\delta=-C_Q/2C_0$. Then the phase of the reflected signal is
\begin{equation}
\label{phshift}
  \theta=\mathrm{tan}^{-1}\left( 2QC_Q/C_0 \right).
\end{equation}

For our measurements, an RF excitation of -119 dBm
($V_{in}\simeq0.2 \mu$V) was reflected from the tank circuit. The
in-phase and quadrature signals from the mixer were low-pass
filtered using a cut-off frequency of 10 kHz. For each value of
$V_\mathrm{SD}$, $V_g$ was ramped at 237 Hz and 1024 repetitions
of the signal were acquired and averaged taking a few seconds.
Below $eV_\mathrm{SD}=4E_C$ only a small modulation of the
magnitude of the reflected RF excitation was observed,
corresponding to $R_{SET}>1$ M$\Omega$. This value of $R_{SET}$ is
much too large to produce the phase shifts observed at these
biases as described above. In Fig. \ref{phase}a we show the phase
shift of the reflected signal as a function of $V_\mathrm{SD}$ and
$n_g$. The phase shifts fall along the Cooper-pair degeneracies
and are concentrated in characteristic ``X'' patterns around the
DJQP points with a maxima at the center. The X's are cut off for
$eV_\mathrm{SD}<E_C$ and for $eV_\mathrm{SD} \gtrsim 3E_C$ This
pattern can be understood by considering the QP tunneling rates
involved in the DJQP cycles and illustrated in Fig. \ref{phase}b.
For gate charge $n_g\in[0,1]$, the charge states 0 and 2 are
degenerate along the negative-slope dotted line in the right half
of Fig. \ref{phase}a. The charge states -1 and 1 are degenerate
along the positive-slope line. Between $eV_\mathrm{SD}=E_C$ and
$eV_\mathrm{SD}=3E_C$, the QP transitions, \textit{e.g} $2
\rightarrow 1$ and $1 \rightarrow 0$ involve an energy gain
$\Delta E<2\Delta$, and as discussed previously, these rates are
small. Nevertheless, away from the DJQP point, the system spends
some fraction of time in a non-degenerate charge state, which
reduces the observed phase shift. One observes a maximum phase
shift precisely at the DJQP, since both the $0 \leftrightarrow 2$
and $-1 \leftrightarrow 1$ manifolds are at their Cooper-pair
charge degeneracies, despite the QP transitions that move the
system slowly from one manifold to the other. For
$eV_\mathrm{SD}<E_C$, the QP transitions $1 \rightarrow 0$ and $0
\rightarrow 1$ involve an energy cost, therefore their rates are
thermally suppressed and the system gets stuck in the
non-degenerate states 1 or 0. For $eV_\mathrm{SD} \gtrsim 3E_C$,
since $E_C \simeq \Delta/2$, the QP transitions $2 \rightarrow 1$
and $-1 \rightarrow 0$ have an energy gain greater than $2\Delta$
and become exponentially larger. Again, the system is trapped in
the non-degenerate states 1 or 0.

In Fig. \ref{phase}c we show the phase shift versus gate charge
for $eV_\mathrm{SD}=2E_C$, \textit{i.e.} crossing the DJQP, and
for $eV_\mathrm{SD} = 1.35E_C$. We can fit the measured phase
shifts using the two-level approximation and Eq. \ref{thermal_cq},
and get good agreement using the values of $E_J$ found from the
spectroscopic measurements discussed below, if we allow for a
non-zero temperature.
%\footnote{We note that at the charge
%degeneracy, the quantum capacitance is reduced by the factor
%tanh$(E_J/2k_BT)$}.
%In this
%case, the phase shift is given by
%\begin{equation}
%\label{ftphshift}
%  \theta=-Q\frac{C_Q}{C_T}\frac{\partial}{\partial q} \frac{q}{\sqrt{q^2+r^2}}
%  tanh\left( \frac{2E_C}{k_BT}\sqrt{q^2+r^2}\right).
%\end{equation}
A second free parameter is an overall constant, which is reduced
from the value $2Q$ in Eq. \ref{phshift} due to imperfections of
our microwave circuitry, \textit{e.g.} a less than ideal
directivity of the directional coupler used in our setup. This
constant is further reduced away from the DJQP point due to the
relative time spent in the non-degenerate manifold. The extracted
temperature is $T=130$ mK, which is higher than the bath
temperature $T\simeq20$ mK. This could be due to self-heating from
the small $\sim$1 pA current produced by the JQP
cycles\footnote{Calculations as described in Verugh \textit{et
al.} J. Appl. Phys. \textbf{78}, 2830 (1995) for self heating in
single-electron devices, using the parameters of our sample,
produce a temperature of ~150mK.}. Another factor contributing to
the elevated temperature could be noise from the cold amplifier,
which can be reduced by using a cold microwave circulator.

\begin{figure}[tb]
    \centering
    \includegraphics[width = 0.5\columnwidth]{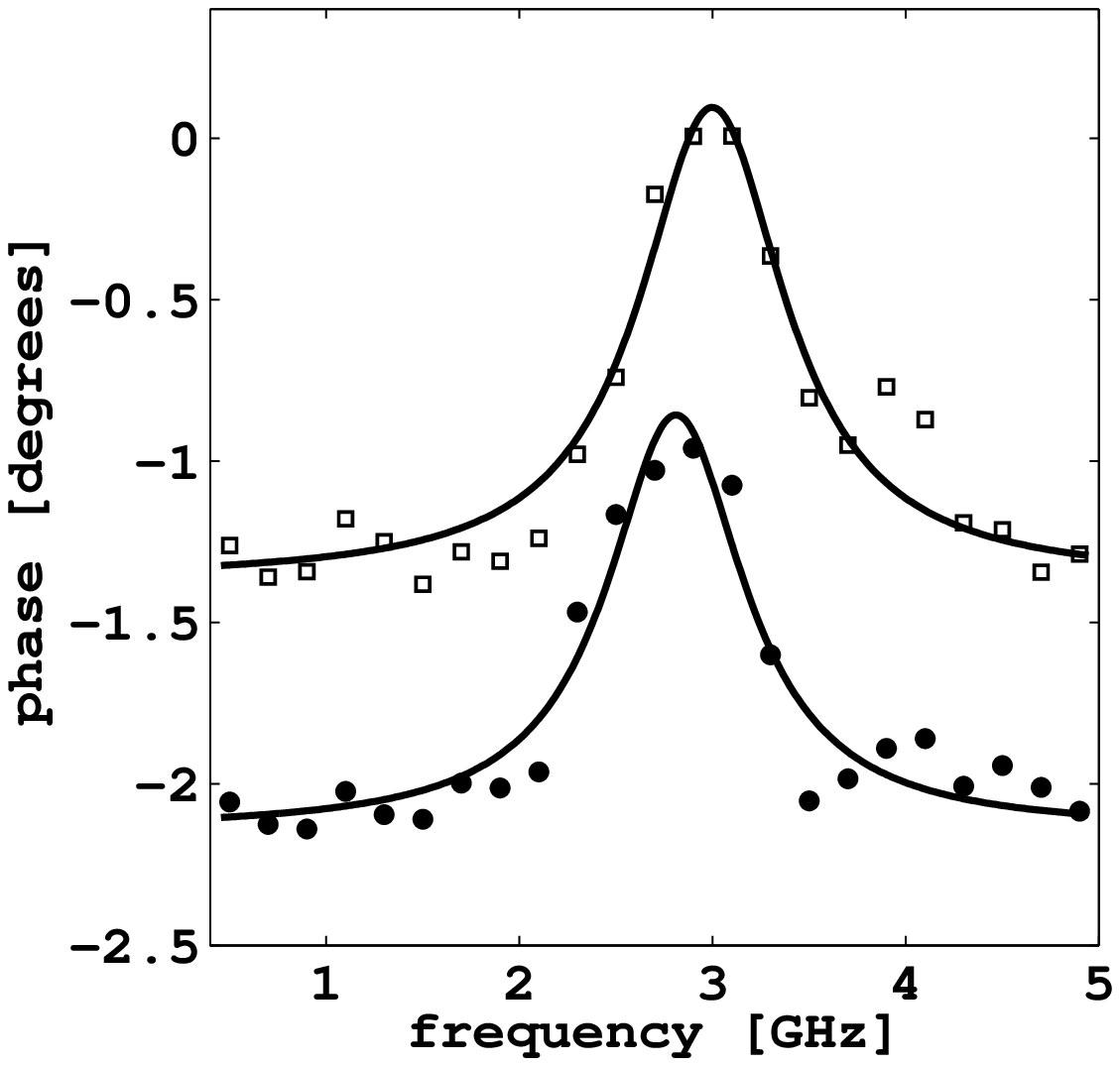}
    \caption{Microwave spectroscopy for $V_\mathrm{SD}=170\mu V$ and
    gate charge $n_g$ tuned to the Cooper-pair resonance of
    junction 1 (upper data) and junction 2 (lower data). The
    lower data set has been offset -1 degrees for clarity. The solid
    lines are fits to a Lorentzian and produce frequencies 3.0 MHz for
    junction 1, and 2.8 MHz for junction 2, and a FWHM of 0.9 MHz for both.
}
    \label{spect}
\end{figure}
We can make a more direct measurement of $E_J$ for each junction
by applying microwaves to the gate of the transistor. Resonant
microwave radiation induces transitions to the excited state,
which has a capacitance of the opposite sign. Fig. \ref{spect}
shows the effects of microwaves on the phase shift when the
transistor is tuned in $V_\mathrm{SD}$ and $n_g$ to be at the
Cooper-pair degeneracy for the individual junctions. The data are
fit to Lorentzians, and the resulting values $E_J/h=3.0$ GHz for
junction 1, and $E_J/h=2.8$ GHz for junction 2 agree well with the
estimate $2.9$ GHz derived from the normal state resistance using
the Ambegaokar-Baratoff relation. The full width at half maximum
(FWHM) of both fits is 0.9 GHz. It is due to lifetime broadening
and indicates a relaxation time $T_1=1.1$ ns, which is consistent
with our previous measurements on charge qubits\cite{duty_prb04},
given that this transistor is more strongly coupled to its
environment.

A suitable device for studying the quantum capacitance in detail
would consist of a SCB placed in a RF-tank circuit. An
electrometer based upon such a RF-SCB device could go beyond the
so-called shot noise limit, which is due to the source-drain
current in the transistor\cite{aassime01b,johansson02}. Moreover,
the RF-SCB would form an integrated charge qubit and readout
device. While this has been done using microwave resonators to
reach the cavity-QED, strong-coupling limit\cite{wallraff04},
%---where the resonator frequency is made to
%match the qubit splitting, and the electromagnetic field of the
%cavity is that of a single photon\cite{wallraff04}. We suggest
we suggest that it is not necessary to go to such an extreme
quantum limit simply for qubit readout. Instead one can us
lower-frequency lumped circuits that utilize the quantum
capacitance, \textit{i.e.} the response of the SCB to classical
electromagnetic fields. Using lower frequency resonators may
further shield the qubit from the high-frequency fluctuations of
the electromagnetic environment.

In conclusion, we have observed the quantum contribution to the
effective capacitance of a Cooper-pair transistor by measuring the
gate-dependent phase shift of a resonant circuit in which the
transistor is embedded. The measured phase shifts are in good
agreement with a theory that takes into account a finite
temperature and the combined tunneling of Cooper-pairs and
quasiparticles.

\begin{acknowledgments}
We would like to acknowledge helpful discussions with V. Shumeiko.
The work was supported by the Swedish SSF and VR, by the
Wallenberg foundation, and by the EU under the IST-SQUBIT-2
programme.
\end{acknowledgments}
\bibliography{qcap_v2}
\end{document}